\documentclass[fleqn,twoside]{article}
\usepackage{espcrc2}

\usepackage{graphicx}
\usepackage[figuresright]{rotating}

\newcommand{\AmS}{{\protect\the\textfont2
  A\kern-.1667em\lower.5ex\hbox{M}\kern-.125emS}}

\title{Analytic Amplitudes for
Hadronic Forward Scattering : COMPETE update}

\author{B. Nicolescu\address[LPNHE]{LPNHE (Unit\'e de Recherche des Universit\'es Paris 6 et Paris 7, Associ\'ee au CNRS), Universit\'e Pierre et Marie Curie, Tour 12 E3, 4 
    Place Jussieu, 75252 Paris Cedex 05, France},
        J. R. Cudell\address{Institut de Physique, B\^at. B5, Universit\'e de 
    Li\`ege, Sart Tilman, B4000 Li\`ege, Belgium},
        V. V. Ezhela\address[COMPAS]{COMPAS group, IHEP, Protvino, Russia},
P. Gauron\addressmark[LPNHE],
K. Kang\address{Physics Department, Brown University, Providence, RI, 
    U.S.A.},
Yu.~V.~Kuyanov\addressmark[COMPAS],
S.~B.~ Lugovsky\addressmark[COMPAS],
E. Martynov\address{Bogolyubov Institute for Theoretical Physics,
03143
Kiev, Ukraine},
E.~ A.~ Razuvaev\addressmark[COMPAS],
N. P. Tkachenko\addressmark[COMPAS]}
       
\begin{document}

\begin{abstract}
      We consider several classes of analytic parametrizations of hadronic 
        scattering amplitudes, and compare their predictions to all available 
        forward data ($pp, \bar pp, \pi p, Kp,$ 
 $\gamma p, \gamma\gamma,\Sigma 
        p$). Although these parametrizations are very close for $\sqrt{s}\ge 
        9$ GeV, it turns out that they differ markedly at low energy, where 
        a universal Pomeron term $\sim\ln^2s$ enables one to extend the fit 
        down to $\sqrt{s}=4$ GeV.
  We present predictions on the total cross sections 
and on the ratio of the real part to the imaginary part of the 
elastic amplitude
($\rho$ parameter) for present and future $pp$ and $\bar p p$
colliders, and on total cross sections 
for $\gamma p \to$ hadrons at cosmic-ray energies and
for $\gamma\gamma\to$ hadrons up to $\sqrt{s}=1$ TeV. 

\vspace{1pc}
\end{abstract}

\maketitle

Analytic parametrizations of forward $(t=0)$ hadron scattering amplitudes is a well-established domain in strong interactions.

However, in the past, the phenomenology of forward scattering hadquite a high degree of arbitrariness :
i) Excessive attention was paid to
$pp$ and $\bar pp$ scattering ; 
ii) Important physical constraints were
often mixed with less general 
    or even ad-hoc properties ;
iii) The
cut-off in energy, defining the region of applicability of the
high-energy models, differed from one author to the other ;
iv) The set of
data considered by different authors was often different ;
v) No rigorous
connection was made between the number of parameters 
    and the number of
data points ;
 vi) No attention was paid to the necessity of the stability
of 
    parameter values ;
vii) The experiments were performed in the past
in quite a chaotic 
    way : huge gaps are sometimes present between
low-energy and 
    high-energy domains or inside the high-energy domain
itself.

The COMPETE (\underline{CO}mputerized \underline{M}odels and
\underline{P}arameter \underline{E}valuation for \underline{T}heo\-ry
and \underline{E}xperiment) collaboration has cured 
as much as possible
the above discussed arbitrariness.

The $\chi^2/dof$ criterium is not
able, 
by itself, to cure the above difficulties : new indicators have 
to
be defined.
Once these indicators are defined\cite{Cudell},  it is possible
to estimate the overall performance of each model, and to establish a
ranking : the highest the numerical value of the rank 
the better the model under consideration.

The final aim of the COMPETE project is to provide our community with a 
periodic cross assessments of data and models via computer-readable 
files on the Web \cite{web}.

We consider the following exemplar cases of the imaginary part of the 
scattering amplitudes :
\begin{equation}
    ImF^{ab}=s\sigma_{ab}=P_{1}^{ab}+P_{2}^{ab}+R_{+}^{ab}
    \pm R_{-}^{ab}
    \label{eq:1prime}
\end{equation}
where : \\
- the $\pm$ sign in formula (\ref{eq:1prime}) corresponds to 
antiparticle (resp. particle) - particle scattering amplitudes.\\
- $R_{\pm}$ signify the effective secondary-Reggeon 
($(f,a_{2}),\ (\rho,\omega)$) contributions to the 
even (odd)-under-crossing amplitude
\begin{equation}
    R_{\pm}(s)=Y_{\pm}\left(\frac{s}{s_{1}}\right)^{\alpha_{\pm}},
    \label{eq:2}
\end{equation}
where $Y$ is a constant residue, $\alpha$ - the reggeon intercept and 
$s_{1}$ - a scale factor fixed at 1 GeV$^2$ ;\\
- $P_{1}(s)$ is the contribution of the Pomeron Regge pole
\begin{equation}    
    P_{1}^{ab}(s)=C_{1}^{ab}\left(\frac{s}{s_{1}}\right)^{\alpha_{P_{1}}},
    \label{eq:3}
\end{equation}
$\alpha_{P_{1}}$ is the Pomeron intercept
$
    \alpha_{P_{1}}=1,
$
and $C^{ab}$ are constant residues.\\
- $P_{2}^{ab}(s)$ is the second component of the Pomeron 
corresponding to three different $J$-plane singularities :
\begin{itemize}
\item[a)] a Regge simple - pole contribution
    \begin{equation}    
    P_{2}^{ab}(s)=C_{2}^{ab}\left(\frac{s}{s_{1}}\right)^{\alpha_{P_{2}}}, 
    \label{eq:5}
    \end{equation}    
with
    $\alpha_{P_{2}}=1+\epsilon,\  \epsilon >0,$
    and $C_{2}^{ab}$
constant ;
     
    \item  [b)] a Regge double-pole contribution
    \begin{equation}                        
        P_{2}^{ab}(s)=s\left[A^{ab}+B^{ab}\ln\left(\frac{s}{s_{1}}
        \right)\right], 
    \label{eq:7}
    \end{equation}      
   with $A^{ab}$ and $B^{ab}$ constant ;
      
    \item  [c)] a Regge triple-pole contribution
    \begin{equation}
        P_{2}^{ab}(s)=s\left[A^{ab}+B^{ab}\ln^2\left(\frac{s}{s_{0}}
        \right)\right],
        \label{eq:8}
    \end{equation}
where $A^{ab}$ and $B^{ab}$ are constants and $s_{0}$ is an arbitrary 
scale factor.
\end{itemize}
We consider all the existing forward data for $pp,\bar pp, \pi p, Kp, 
\gamma\gamma$ and $\Sigma p$ scatterings.
The number of data points is : 904, 742, 648,  569, 498, 453, 397, 
329 when the cut-off in energy is 3, 4, 5, 6, 7, 8, 9, 10 GeV 
respectively.
A large number of variants were studied. 
All definitions and numerical details can be found in Ref. 1.

The 2-component Pomeron classes of models are
    RRPE, RRPL and RRPL2,
where by RR we denote the two effective secondary-reggeon 
contributions, by P - the contribution of the Pomeron Regge-pole 
located at $J=1$, by E - the contribution of the Pomeron Regge-pole 
located at $J=1+\epsilon$, by L - the contribution of the component of 
the Pomeron, located at $J=1$ (double pole), and by L2 - the contribution of the 
component of the Pomeron located at $J=1$ (triple pole).
We also studied the 1-component Pomeron classes of models
    RRE, RRL and RRL2. 
  
The highest rank are get by the RRPL2$_u$ mo\-dels (see Table 1), corresponding to the $\ln^2s$ be-

\begin{table}[h]
\caption{Ranking of the the 21 models having nonzero area of applicability.
The number between paranthesis, in the Model Code column, denotes the number of free parameters.}
{\centering \begin{tabular}{lr}
\hline
 Model Code&
 Rank \\
\hline
 \( {\textrm{RRPL}2_{u}(19)}\)&
 230 \\
\hline
 \( {\textrm{RRP$_{nf}$L}2_{u}(21)} \)&
 222 \\
\hline
\( {\textrm{RRL}_{nf}(19)} \)&
 222 \\
\hline
 \( {(\textrm{RR}_{c})^{d}\textrm{ PL}2_{u}}(15) \)&
 204 \\
\hline
 \( {(\textrm{RR})^{d}\textrm{ P$_{nf}$L}2_{u}}(19) \)&
 194 \\
\hline
 \( {[\textrm{R}^{qc}\textrm{ L}^{qc}]\textrm{R}_{c}}(12) \)&
184 \\
\hline
 \( {(\textrm{RR}_{c})^{d}\textrm{P}^{qc}\textrm{L}2_{u}}(14) \)&
 181 \\
\hline
 \( {(\textrm{RR})^{d}\textrm{P}^{qc}\textrm{L}2_{u}}(16) \)&
 180 \\
\hline
 \( {\textrm{RR}_{c}\textrm{ L}2^{qc}(15)} \)&
180 \\
\hline
 \( {(\textrm{RR})^{d}\textrm{ P$_{nf}$L}2(20)} \)&
178 \\
\hline
 \( {(\textrm{RR})^{d}\textrm{ PL}2_{u}}(17) \)&
174 \\
\hline
 \( {\textrm{RRPL}(21)} \)&
173 \\
\hline
 \( {\textrm{RR}_{c}\textrm{ L}^{qc}}(15) \)&
 172 \\
\hline
 \( {\textrm{RRL}2^{qc}(17)} \)&
170 \\
\hline
 \( {[\textrm{R}^{qc}\textrm{ L}2^{qc}]\textrm{R}_{c}}(12) \)&
 170 \\
\hline
 \( {\textrm{RRL}^{qc}(17)} \)&
162 \\
\hline
 \( {\textrm{RRPE}_{u}}(19) \)&
 158 \\
\hline
 \( {[\textrm{R}^{qc}\textrm{L}^{qc}]\textrm{R}}(14) \)&
155 \\
\hline
 \( {\textrm{RRL}2(18)} \)&
 152 \\
\hline
 \( {\textrm{RR}_{c}\textrm{PL}}(19) \)&
142 \\
\hline
 \( {\textrm{RRL}(18)} \)&
 133\\
\hline
\end{tabular}\par}
\label{Table1}
\end{table}

\begin{table}[h]
\caption{Predictions for $\sigma_{tot}$ and $\rho$, for $\bar pp$ (at
$\sqrt{s}=1960$ GeV) and for $pp$ (all other energies). The central values 
and statistical errors correspond to the preferred model RRPL2$_u$.}
{\centering \begin{tabular}{ccc}
\( \sqrt{s} \) (GeV)& \( \sigma \) (mb)& \( \rho \)\\
\hline
100& \( 46.37\pm 0.06\)& 
 \( 0.1058\pm 0.0012\)\\
200& \( 51.76\pm 0.12 \)& 
 \( 0.1275\pm 0.0015\)\\
300& \( 55.50\pm 0.17\)&
 \( 0.1352\pm 0.0016\)\\
400& \( 58.41\pm 0.21 \)&
 \( 0.1391\pm 0.0017\)\\
500& \( 60.82\pm 0.25 \)& 
 \( 0.1413\pm 0.0017\)\\
600& \( 62.87\pm 0.28 \)&
 \( 0.1416\pm 0.0018\)\\
1960& \( 78.27\pm 0.55 \)&
 \( 0.1450\pm 0.0018\)\\
10000& \( 105.1\pm 1.1 \)&
 \( 0.1382\pm 0.0016\)\\
12000& \( 108.5\pm 1.2 \)& 
 \( 0.1371\pm 0.0015\)\\
14000& \( 111.5\pm 1.2\)& 
 \( 0.1361\pm 0.0015\)\\
\end{tabular}\par}
\label{table2}
 \end{table}

\begin{table}[h]
\caption{Predictions for $\sigma_{tot}$ for $\gamma p \to hadrons$ for 
cosmic-ray photons. The central values and the statistical errors are as in Table \ref{table2}.}
{\centering \begin{tabular}{cc}
\( p_{lab}^{\gamma} \) (GeV)& \( \sigma \) (mb)\\
\hline
$0.5\cdot10^6$&\( 0.243\pm 0.009\)\\
$1.0\cdot10^6$&\( 0.262\pm 0.010\)\\
$0.5\cdot10^7$&\( 0.311\pm 0.014\)\\
$1.0\cdot10^7$&\( 0.333\pm 0.016\)\\
$1.0\cdot10^8$&\( 0.418\pm 0.022\)\\
$1.0\cdot10^9$&\( 0.516\pm 0.029\)\\
\end{tabular}\par}
\label{table3}
\end{table}
\noindent haviour of total cross sections first proposed by Heisenberg 50 years ago \cite{Heisen52}.
The $u$ index denotes the \textit{universality property} :  
the coupling $B$ of the $\ln^2s$ term is the same in all hadron-hadron scatterings and $s_0$ is the same in all reactions.

\begin{table}[h]
\caption{Predictions for $\sigma_{tot}$ for $\gamma \gamma \to hadrons$. 
The central values and the statistical errors  are as in Table \ref{table2}.}
{\centering \begin{tabular}{cc}
\( \sqrt{s} \) (GeV)& \( \sigma \) ($\mu$ b)\\
\hline
200&\( 0.546\pm 0.027\)\\
300&\( 0.610\pm 0.035 \)\\
400&\( 0.659\pm 0.042 \)\\
500&\( 0.700\pm 0.047 \)\\
1000&\( 0.840\pm 0.067 \)\\
\end{tabular}\par}
\label{table4}
\end{table}

We note that the familiar RRE Donnachie-Landshoff model  is \textit{rejected} at the 98\% C.L. when models which achieve a $\chi^2/dof$ less than 1 for $\sqrt{s}\ge 5$ GeV are considered. 

The predictions of the best RRPL2$_{u}$
model, adjusted for $\sqrt{s} \geq 5$ GeV,  are given in Tables 2-4. 

The uncertainties on total cross sections, including the systematic errors 
due to contradictory data points from FNAL (the CDF and E710/E811 experiments, respectively),
can reach $1.9\%$ at RHIC, $3.1\%$ at the Tevatron, and $4.8\%$ at the LHC,
whereas
those on the $\rho$ parameter are respectively $5.4\%$, $5.2\%$, and $5.4\%$.
The global picture emerging from fits to all data on forward observables
supports 
the CDF data and disfavors the 
E710/E811 data at $\sqrt{s}=1.8$ TeV.

Any significant deviation from the predictions based on model RRPL2$_u$ 
will lead to a re-evaluation of the hierarchy of models and presumably
change the preferred parametrisation to another one. 
A deviation from the ``allowed region" 
would be an indication that strong interactions demand
a generalization of the analytic models discussed so far, {\it e.g.} by adding
Odderon terms, or new Pomeron terms, as suggested by QCD.


\begin{thebibliography}{99}
    \bibitem{Cudell}  J. R. Cudell, V. V. Ezhela, P. Gauron, K.~Kang, Yu. 
    V. Kuyanov, S. B. Lugovsky, B.~Nicolescu, and N.~P.~Tkachenko, 
    Phys. Rev. \textbf{D65} (2002) 074024 ; see also 2002 Review of Particle Physics, K. Hagiwara et al. Phys. Rev. \textbf{D66} (2002) 010001-9.

\bibitem{web}  See the preliminary version of a Web interface at 
    the adress http://\\ sirius.ihep.su/$^\sim$kuyanov/OK/eng/intro.html. 
    
 \bibitem{Heisen52}  W. Heisenberg, Zeit. Phys. \textbf{133} (1952) 65 
   (in German).
 \bibitem{JR}J.~R.~Cudell, V.~V.~Ezhela, P.~Gauron, K.~Kang, 
Yu.~V.~Kuyanov, S.~B.~Lugovsky, E. Martynov, B.~Nicolescu, E.~A.~Razuvaev, and N.~P.~Tkachenko, hep-ph/0206172, Phys. Rev. Lett. (in press).
\end{thebibliography}
\end{document}